\documentclass[amsmath,floatfix,twocolumn,superscriptaddress]{revtex4-1}
\usepackage{amssymb,amsmath}
\usepackage{graphicx,array}
\usepackage{dcolumn}
\usepackage{psfrag}
\usepackage{bm}
\usepackage{color}

\begin{document}
\title{Giant Thermal Enhancement of the Electric Polarization in Ferrimagnetic 
       BiFe$_{1-x}$Co$_{x}$O$_{3}$ Solid Solutions Near Room Temperature}

\author{C\'{e}sar Men\'{e}ndez}
\author{Claudio Cazorla}
\affiliation{School of Materials Science and Engineering, UNSW Sydney, Sydney, NSW 2052, Australia}

\begin{abstract}
Thermal excitations typically reduce the electric polarization in ferroelectric materials. Here, we 
show by means of first-principles calculations that multiferroic BiFe$_{1-x}$Co$_{x}$O$_{3}$ solid 
solutions with $0.25 \le x \le 0.50$ (BFCO) represent a noteworthy exception to this behaviour. 
In particular, we find that at room temperature and for moderate pressures of $0.1$--$1.0$~GPa, 
depending on the composition, the electric polarization of bulk BFCO increases by $\sim 200$\%. 
The origin of such an exceptional behavior is a phase transformation involving a low-$T$ rhombohedral 
(${\cal R}$) phase and a high-$T$ super-tetragonal (${\cal T}$) phase. Both ${\cal R}$ and ${\cal T}$ 
phases are ferrimagnetic near room temperature with an approximate net magnetization of $0.13$~$\mu_{B}$ 
per formula unit. Contrarily to what occurs in either bulk BiFeO$_{3}$ or BiCoO$_{3}$, the ${\cal T}$ 
phase is stabilized over the ${\cal R}$ by increasing temperature due to its higher vibrational entropy. 
This extraordinary $T$-induced ${\cal R} \to {\cal T}$ phase transition is originated by polar phonon 
modes involving concerted displacements of transition-metal and oxygen ions.
\end{abstract}
\maketitle

Super-tetragonal (${\cal T}$) oxide perovskites comprise a family of materials that are fundamentally
intriguing and have great potential for ferroelectric, piezoelectric, sensor, and energy conversion applications 
\cite{zhang18,yamada13,infante11}. Super-tetragonal phases exhibit giant electric polarizations of the order 
of $100$~$\mu$C/cm$^{2}$ and may be accompanied by magnetism \cite{chen12,kuo16,park14}. The coexistence of 
ferroelectricity and magnetism in crystals, known as multiferroics, offers the possibility of controlling the 
magnetization with electric fields via their order-parameter coupling. Such a magnetoelectric coupling can be 
used, for example, to design ultra efficient logic and memory devices and realize large piezomagnetic coefficients 
for the miniaturization of antennas and sensors \cite{heron14,allibe12,domann17,nan17}. Furthermore, phase transitions 
involving ${\cal T}$ phases typically exhibit colossal volume changes of $\sim 10$\% (e.g., PbVO$_{3}$ and related 
solid solutions), which can be exploited in mechanical degradation \cite{yamamoto19,pan19} and solid-state cooling 
\cite{manosa17,cazorla19,cazorla20} applications. Examples of ${\cal T}$ multiferroic materials are bulk BiCoO$_{3}$ 
(BCO) and BiFeO$_{3}$ (BFO) thin films \cite{belik06,wang03}.

Nonetheless, ${\cal T}$ phases usually are thermodynamically too stable and hence difficult to switch by means of an 
external field or temperature, which severely limits their technological applicability. For example, in order to 
stabilize a paraelectric phase in multiferroic ${\cal T}$ BiCoO$_{3}$ it is necessary to increase its temperature 
above $800$~K or apply a large hydrostatic pressure of $P > 3$~GPa \cite{oka10,cazorla17,cazorla18}. 
Likewise, the region in which the functionality of super-tetragonal BiFeO$_{3}$ thin films can be exploited corresponds 
to a narrow epitaxial strain interval in which the ${\cal T}$ phase coexists with a different polymorph and as a result 
becomes structurally soft \cite{heo17,iniguez10}. Moreover, ${\cal T}$ multiferroics mostly are antiferromagnetic 
(i.e., their atomic magnetic moments align antiparallel rendering negligible net magnetizations) and consequently 
are unresponsive to external magnetic fields \cite{macdougall12}. Therefore, it is highly desirable to find new 
${\cal T}$ multiferroic materials that react significantly to external bias near ambient conditions. 

\begin{figure}[t]
\centerline{
\includegraphics[width=1.00\linewidth]{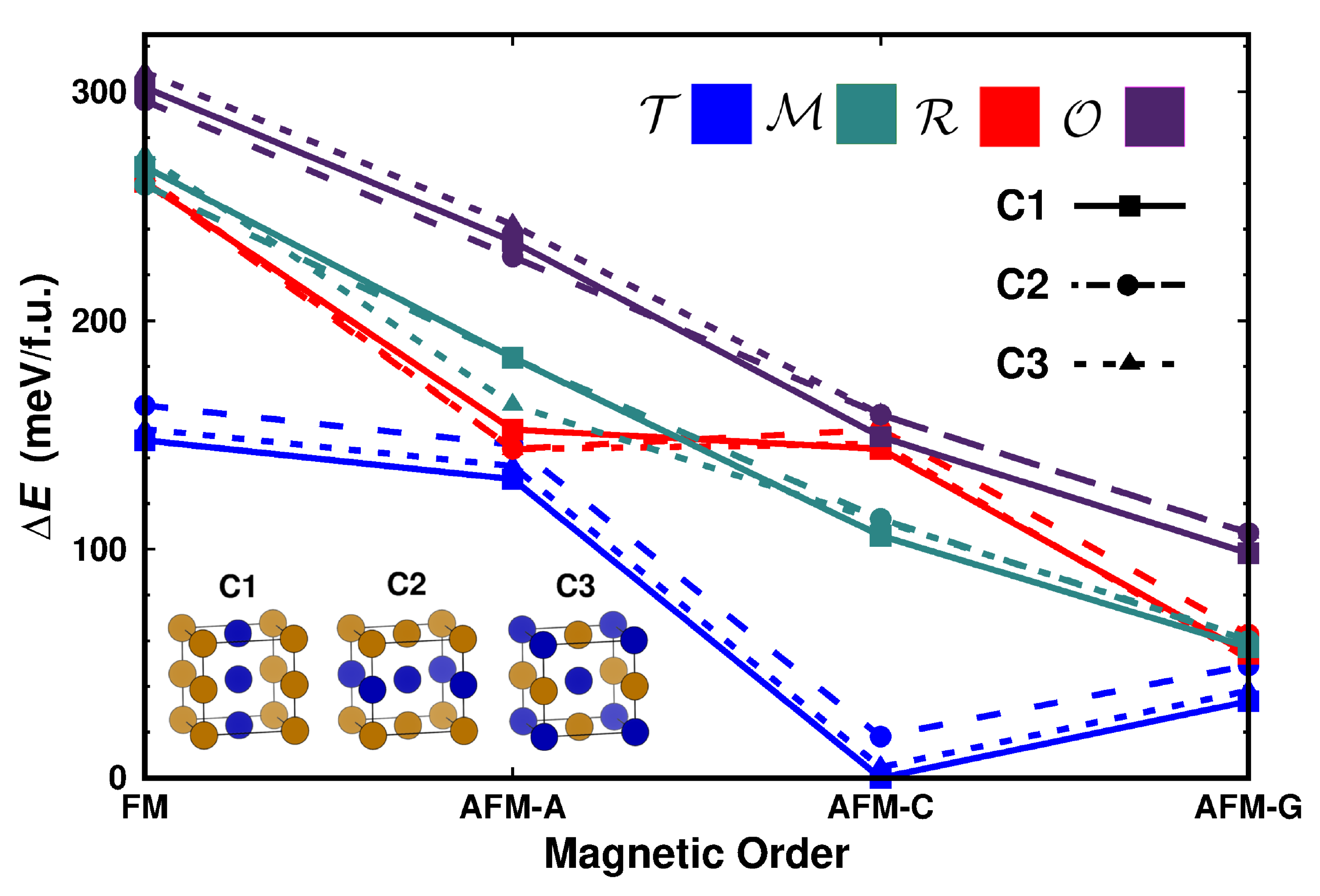}}
\caption{First-principles analysis of bulk BiFe$_{0.5}$Co$_{0.5}$O$_{3}$ at zero pressure and $T = 0$. Crystal structures 
	with tetragonal (${\cal T}$), monoclinic (${\cal M}$), rhombohedral (${\cal R}$), and orthorhombic (${\cal O}$) 
	symmetry were considered (Supplementary Fig.1). All possible Co--Fe (C1, C2, and C3) and magnetic spin arrangements 
	(ferromagnetic --FM-- and antiferromagnetic --AFM-- of type A, C, and G --Supplementary Fig.1--) were generated for a 
	$2 \times \sqrt{2} \times \sqrt{2}$ simulation cell containing $20$ atoms \cite{cazorla15}.}
\label{fig1}
\end{figure}

In this Letter, we show by means of first-principles calculations based on density functional theory (DFT) 
that BiFe$_{1-x}$Co$_{x}$O$_{3}$ solid solutions (BFCO) with $0.25 \le x \le 0.50$ represent ideal bulk systems 
in which to realize the full potential of multiferroic ${\cal T}$ phases. Specifically, we find that under 
moderate hydrostatic pressures of $0.1 \lesssim P \lesssim 1$~GPa (depending on the composition) it is possible 
to trigger a phase transition from a low-$T$ rhombohedral (${\cal R}$) phase to a high-$T$ ${\cal T}$ phase 
at room temperature. The disclosed $T$-induced ${\cal R} \to {\cal T}$ phase transformation involves (i)~a 
colossal increase in the electric polarization of $\Delta p \sim 200$\%, (ii)~the existence of a robust 
net magnetization of $\approx 0.13$~$\mu_{B}$ per formula unit (f.u.), and (iii)~a giant volume increase of 
$\Delta V \sim 10$\%. Examples of technologies in which these multifunctional phenomena could have an immediate 
impact include pyroelectric energy harvesting (the $T$-induced variation of the electric polarization is 
tremendous \cite{bowen14,hoffmann15}) and solid-state cooling (multicaloric effects involving the materials 
response to both pressure and magnetic fields could be engineered to overcome practical limiting issues 
\cite{gottschall18}). Meanwhile, the appearance of ferrimagnetism and observation that temperature 
stabilizes the ${\cal T}$ phase over the ${\cal R}$, effects that are missing in bulk BiCoO$_{3}$ and BiFeO$_{3}$, 
pose a series of interesting fundamental questions: Which atomistic mechanisms are responsible for such an 
anomalous $dp / dT \gg 0$ behaviour? What type of thermal excitations drive the uncovered ${\cal R} 
\to {\cal T}$ transformation? Why Co--Fe cation mixing triggers a net magnetization in bulk BFCO? Based on our 
DFT outcomes and analysis, we address these questions and make insightful connections with the experimental 
results reported recently for BiFe$_{1-x}$Co$_{x}$O$_{3}$ solid solutions \cite{azuma08,hojo18,gao18}.

\begin{figure*}[t]
\centerline{
\includegraphics[width=0.90\linewidth]{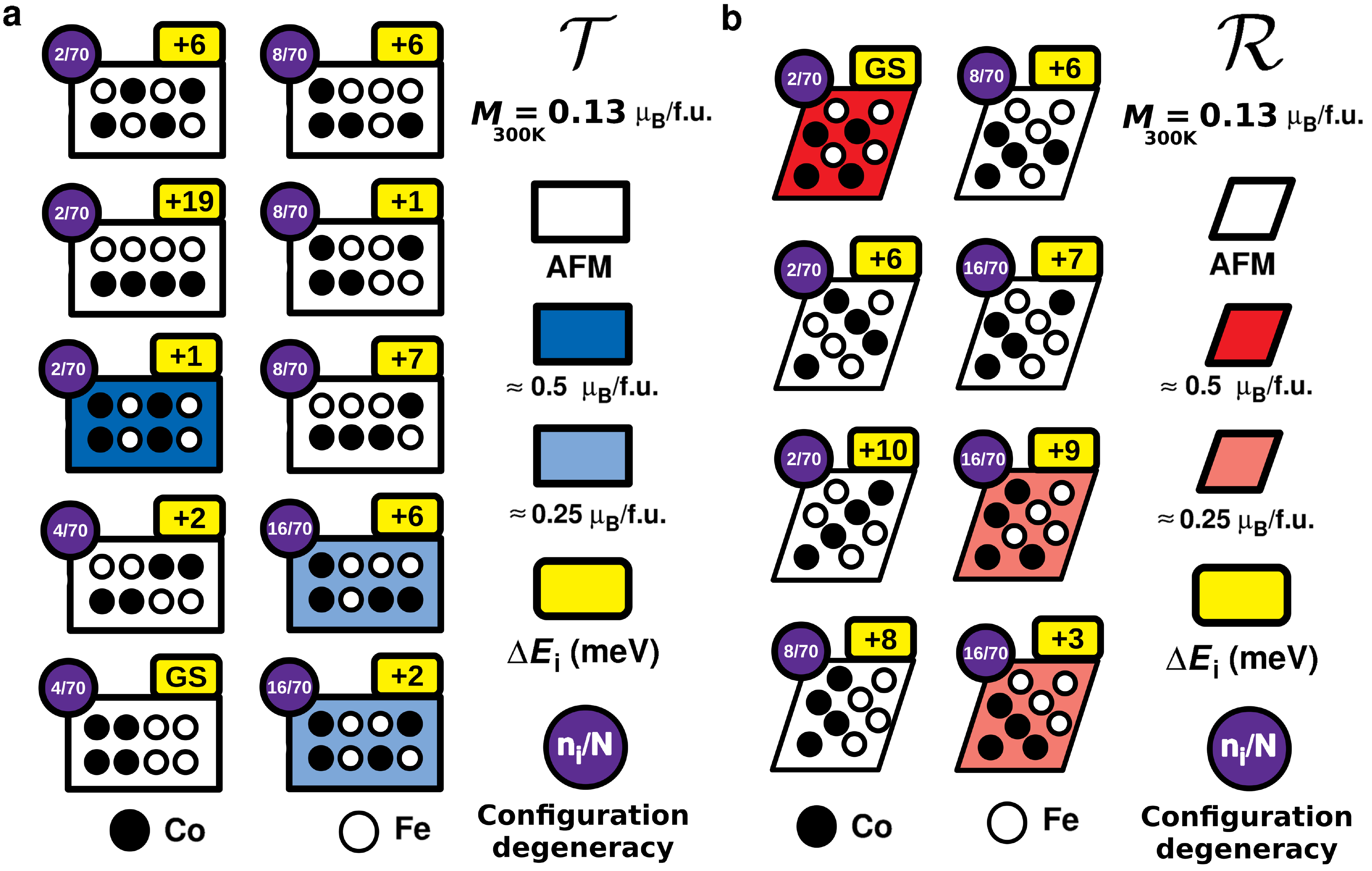}}
\caption{First-principles determination of the magnetic properties of bulk BFCO$_{0.5}$ at finite temperatures
	for phases ({\bf a}) ${\cal T}$ and ({\bf b}) ${\cal R}$. All possible atomic Co--Fe arrangements were 
	generated for a $2 \times 2\sqrt{2} \times \sqrt{2}$ simulation cell containing $40$ atoms, which were reduced 
	by crystal symmetry operations to $10$ ${\cal T}$ and $8$ ${\cal R}$ representative configurations \cite{cazorla19b}. 
	The magnetic moment, total energy, and relative degeneracy of each representative configuration were calculated. 
	The net magnetization estimated for each phase at room temperature is $\approx 0.13$~$\mu_{B}$ per formula unit; 
	``GS'' stands for ground state and $\Delta E_{i} \equiv E_{i} - E_{\rm GS}$.}
\label{fig2}
\end{figure*}

Spin-polarized DFT calculations were performed with the generalized gradient approximation proposed by Perdew, 
Burke and Ernzerhof (PBE) as implemented in the VASP package \cite{vasp,pbe96}. The ``Hubbard-$U$'' scheme 
derived by Dudarev \textit{et al.} was employed for the description of Co~(Fe) $3d$ electrons by adopting a $U$ 
value of $6$~($4$)~eV \cite{cazorla17,cazorla18,cazorla13}. The ``projected augmented wave'' method \cite{bloch94} 
was used to represent the ionic cores by considering the following electronic states as valence: Co $4s^{1}3d^{8}$, 
Fe $3p^{6}4s^{1}3d^{7}$, Bi $6s^{2}5d^{10}6p^{3}$, and O $2s^{2}2p^{4}$. An energy cut-off of $800$~eV and a 
$\Gamma$-centered ${\bf k}$-point grid of $4 \times 6 \times 6$ were employed for a $2 \times \sqrt{2} \times 
\sqrt{2}$ simulation cell containing $20$ atoms \cite{cazorla15}, thus obtaining zero-temperature energies converged 
to within $0.5$~meV/f.u. Geometry relaxations were performed for an atomic force threshold of $0.005$~eV$\cdot$\AA$^{-1}$. 
Electric polarizations were estimated  perturbatively by considering the atomic displacements referred to 
a non-polar reference phase and the corresponding Born effective charges tensor \cite{cazorla15}. {\em Ab initio} 
free energies were calculated within the quasi-harmonic (QH) approximation \cite{cazorla13,cazorla17d} as a function 
of $P$ and $T$. Phonon frequencies were calculated with the small displacement method \cite{kresse95,alfe09}. The following 
technical parameters provided QH free energies converged to within $5$~meV/f.u.: $160$-atom supercells, atomic 
displacements of $0.01$~\AA, and ${\rm q}$-point grids of $16 \times 16 \times 16$ for integration within the first Brillouin 
zone. The effects of chemical disorder were addressed by generating all possible atomic Co--Fe and magnetic spin arrangements 
(ferromagnetic --FM-- and antiferromagnetic --AFM-- of type A, C, and G --Supplementary Fig.1--) for a $2 \times 2\sqrt{2} 
\times \sqrt{2}$ supercell containing $40$ atoms. Quasi-harmonic free energies were calculated only for the lowest-energy 
configurations. Our spin-polarized DFT calculations were performed for bulk BiFe$_{0.5}$Co$_{0.5}$O$_{3}$ and 
BiFe$_{0.75}$Co$_{0.25}$O$_{3}$, hereafter referred to as BFCO$_{0.5}$ and BFCO$_{0.25}$. 

Following a previous work by Di\'eguez and \'I\~niguez \cite{dieguez11}, we considered the four BFCO$_{0.5}$ crystal 
structures that are energetically most competitive at zero temperature. The crystal symmetry of such phases prior 
to introducing chemical disorder on the metal cation sites were tetragonal ($P4mm$), orthorhombic ($Pnma$), monoclinic 
($Pc$), and rhombohedral ($R3c$). The optimized BFCO$_{0.5}$ structures resulting from such parent phases were labelled 
as ${\cal T}$, ${\cal O}$, ${\cal M}$, and ${\cal R}$, respectively (Supplementary Fig.1). Initially, a $20$-atoms unit 
cell was employed to model all four polymorphs and to determine the atomic Co--Fe and magnetic spin arrangements (ferromagnetic 
--FM-- and antiferromagnetic --AFM-- of type A, C, and G --Supplementary Fig.1--) rendering the lowest energy for each 
phase (Fig.\ref{fig1}). The BFCO$_{0.5}$ ground-state phase was identified as ${\cal T}$ with ``C1'' Co--Fe and AFM-C spin 
orderings (Fig.\ref{fig1}) and an electric polarization of $140$~$\mu$C/cm$^{2}$. The first metastable phase lies 
$\sim 45$~meV/f.u. above the ground state and corresponds to a ${\cal R}$ structure presenting ``C3'' Co--Fe and AFM-G 
spin orderings (Fig.\ref{fig1}) and an electric polarization of $45$~$\mu$C/cm$^{2}$. It is worth noting that the energies 
of the ${\cal M}$ and ${\cal R}$ phases are practically degenerate in the AFM-G case. By using analogous computational 
methods to ours, Di\'eguez and \'I\~niguez \cite{dieguez11} concluded that the BFCO$_{0.5}$ ground state was a ${\cal R}$ 
phase with AFM-G spin ordering. The reason for the discrepancy with our results lies on the fact that the DFT exchange-correlation 
functionals employed in both studies are different (Supplementary Fig.2). Nevertheless, our zero-temperature results appear 
to be more consistent with the recent observations by Azuma {\emph et al.}, in which the stable phase of BFCO$_{0.5}$ 
has been experimentally identified as ${\cal T}$ \cite{azuma08,hojo18}. 

To correctly describe the magnetic properties of the ${\cal T}$ and ${\cal R}$ phases at finite temperatures, we 
performed a systematic configurational analysis for a larger simulation cell containing $40$ atoms (Fig.\ref{fig2}). 
The ${\cal T}$ and ${\cal R}$ configurations were initialized with AFM-C and AFM-G spin orderings, respectively, thus 
rendering zero net magnetizations. Upon full optimization, however, a considerable fraction of states exhibited a net 
magnetization of either $0.5$ or $0.25$~$\mu_{B}$/f.u. (Fig.\ref{fig2}) due to spin imbalance between the Co and Fe 
sublattices (Supplementary Fig.3). The ${\cal T}$ phase with the lowest energy displayed AFM-C spin ordering and zero 
net magnetization, while the ${\cal R}$ ground state was ferrimagnetic (FiM) and presented a net magnetization of 
$0.50$~$\mu_{B}$/f.u. (Fig.\ref{fig2}). At $T \neq 0$ conditions, each configuration contributes to the total 
magnetization according to the formula:
\begin{equation}
	M (T) = \sum_{i}^{N_{\rm conf}} M_{i} \cdot \frac{\exp{\left(-\Delta E_{i} / k_{B}T\right)}}{Z_{\rm conf}}~,
\label{eq:magnet}
\end{equation}
where $N_{\rm conf}$ is the total number of configurations, $M_{i}$ ($E_{i}$) the magnetization (energy) of the $i$th 
configuration, $\Delta E_{i} \equiv E_{i} - E_{0}$, $E_{0}$ the ground-state energy, $k_{B}$ the Boltzmann constant, and 
$Z_{\rm conf} \equiv \sum_{i}^{N_{\rm conf}} \exp{\left(-\Delta E_{i} / k_{B}T\right)}$ the configurational partition 
function. By using Eq.(\ref{eq:magnet}) and the energy and configuration degeneracy data reported in Fig.\ref{fig2}, we 
estimated that the net magnetization of the ${\cal T}$ and ${\cal R}$ phases amount both to $0.13$~$\mu_{B}$/f.u. near 
room temperature (Supplementary Fig.4). This result is consistent with a recent work by Gao {\emph et al.} \cite{gao18}, 
in which robust ferrimagnetism has been experimentally demonstrated for BFCO$_{0.5}$ thin films at room temperature.

\begin{figure*}[t]
\centerline{
\includegraphics[width=0.90\linewidth]{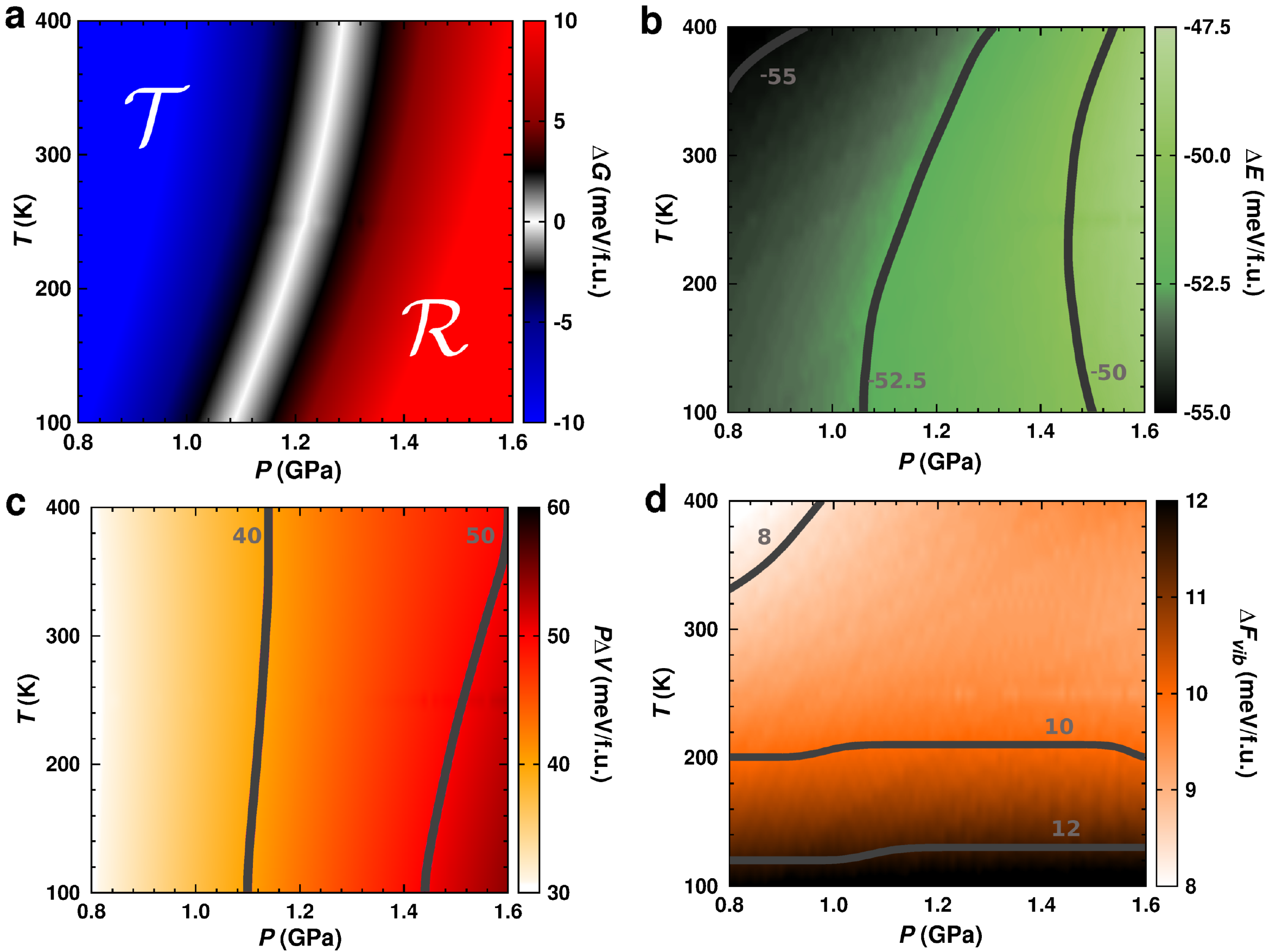}}
\caption{Gibbs free energy difference between the ${\cal T}$ and ${\cal R}$ phases of bulk BiFe$_{0.5}$Co$_{0.5}$O$_{3}$ 
	and their contributions expressed as a function of temperature and pressure ($\Delta A \equiv A_{\cal T} - A_{\cal R}$). 
	{\bf a} Total Gibbs free energy difference. {\bf b} Static internal energy difference. {\bf c} Enthalpy-related energy 
	difference. {\bf d} Vibrational Helmholtz free energy difference. Thermodynamic states presenting equal $\Delta A$ values
	are joined by thick solid lines.}
\label{fig3}
\end{figure*}

Figure~\ref{fig3}a shows the $P$--$T$ phase diagram estimated for bulk BFCO$_{0.5}$ with first-principles methods and 
the QH approximation \cite{cazorla13,cazorla17d}. At low temperatures, a $P$-induced ${\cal T} \to {\cal R}$ phase 
transition occurs around $1$~GPa that is characterized by a huge volume collapse of $\sim 10$\% ($\Delta V = V_{\cal T} 
- V_{\cal R} > 0$), thus indicating a marked first-order behaviour. The corresponding $P$--$T$ phase boundary, determined 
with the condition $\Delta G = G_{\cal T} - G_{\cal R} = 0$ where $G \equiv E + PV - TS$ represents the Gibbs free 
energy and $S$ the entropy, presents a positive slope. Consequently, by the Clausius-Clapeyron relation $\Delta S / \Delta V 
= \partial P / \partial T$, the entropy of the ${\cal T}$ phase should be larger than that of the ${\cal R}$ phase, 
namely, $\Delta S = S_{\cal T} - S_{\cal R} > 0$. Our QH free-energy calculations explicitly confirm this result since 
the value of the Helmholtz free energy difference, $\Delta F = F_{\cal T} - F_{\cal R}$, decreases under uncreasing 
temperature (Fig.\ref{fig3}d) and $\Delta S \equiv - \partial \Delta F / \partial T$. It is noted that the only type 
of entropy considered in our simulations is vibrational, hence the subscript ``vib'' in Fig.\ref{fig3}d, since we 
assume that the magnetic and configurational contributions to $S$ are small at low temperatures and very similar for 
the two phases (i.e., the $\lbrace M_{i} \rbrace$, $\lbrace E_{i} \rbrace$, and configurational degeneracy spectra 
calculated for ${\cal T}$ and ${\cal R}$ are much alike --Fig.\ref{fig2}-- \cite{cazorla19b}) hence they hardly have 
any influence on $\Delta G$.

The BFCO$_{0.5}$ phase diagram shown in Fig.\ref{fig3}a describes an unusual $T$-induced ${\cal R} \to {\cal T}$ 
phase transition occurring at room temperature and a moderate hydrostatic pressure of $1.2 \pm 0.2$~GPa, in which 
the electric polarization of the bulk material increases by $\sim 200$\% (Supplementary Fig.4). Such a $T$-induced
phase transformation involves two different ferroelectric FiM states and is mainly driven by the entropy contributions 
to their Gibbs free energy difference (that is, the $\Delta F$ term). This conclusion is deduced straightforwardly 
from Figs.\ref{fig3}b-d, in which for a fixed $P$ it is observed that the $\Delta E$ and $P \Delta V$ energy differences 
remain practically constant as a function of $T$ (isovalue lines in Figs.\ref{fig3}b,c are practically vertical), 
in marked contrast with $\Delta F$ (isovalue lines in Fig.\ref{fig3}d are practically horizontal). It is worth 
noting that the $T$-induced ${\cal R} \to {\cal T}$ phase transition disclosed here for BFCO$_{0.5}$ has neither 
been predicted nor observed previously in bulk BiCoO$_{3}$ or BiFeO$_{3}$ thin films. For instance, the $P$--$T$ phase 
boundary involving the super-tetragonal phase in bulk BCO presents a negative slope \cite{oka10} due to the presence 
of stiff vibrational modes that reduce the vibrational entropy of the ${\cal T}$ phase as compared to that of other 
competing states \cite{cazorla17,cazorla18}. Consequently, by increasing temperature the stability of the ${\cal T}$ 
phase in bulk BCO is always reduced and the variation of the electric polarization is negative ($dp / dT \ll 0$). A 
very similar behaviour has been found also for bulk ${\cal T}$ BFO \cite{cazorla13}.

\begin{figure*}[t]
\centerline{
\includegraphics[width=0.90\linewidth]{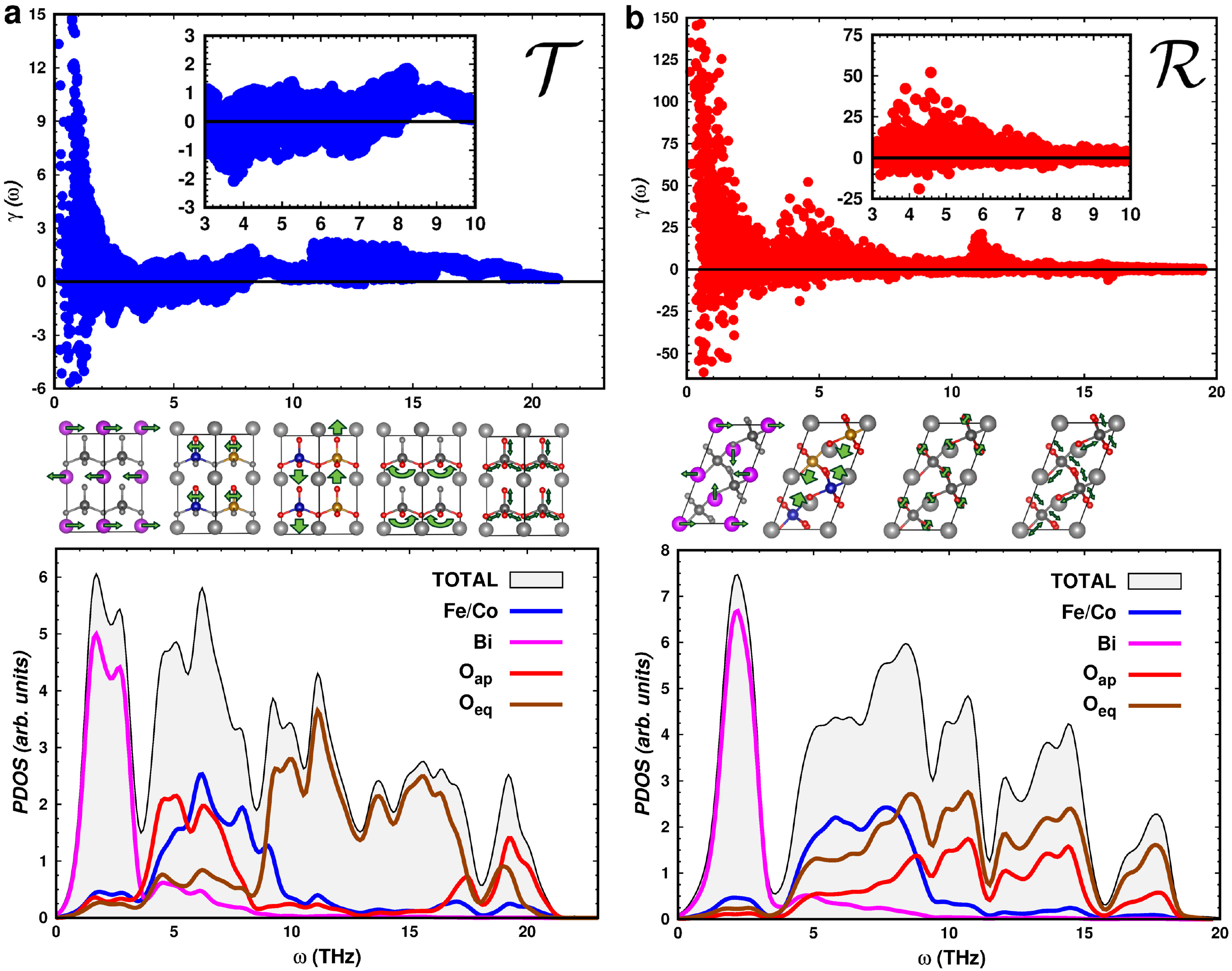}}
\caption{Vibrational properties of the ({\bf a}) ${\cal T}$ and ({\bf b}) ${\cal R}$ phases of bulk 
	 BiFe$_{0.5}$Co$_{0.5}$O$_{3}$. The represented quantities are the Gr\"{u}neisen parameter, 
	 $\gamma (\omega)$, and the density of vibrational states along with the corresponding ionic 
	 contributions (PDOS). O$_{\rm ap}$ and  O$_{\rm eq}$ stand for oxygen atoms in apical and 
	 equatorial positions, respectively. Some representative phonon eigenmodes are sketched with
	 green arrows and they are ordered according to their vibrational frequency.}
\label{fig4}
\end{figure*}

In view of the prominent role played by the lattice excitations on the anomalous $T$-induced stabilization 
of the ${\cal T}$ phase at room temperature, we performed a detailed analysis on the phonon modes and frequencies 
of BFCO$_{0.5}$ (Fig.\ref{fig4}). In particular, we estimated the projected density of vibrational states (PDOS) 
and Gr\"{u}neisen parameter, defined as $\gamma_{i} \equiv - d \ln{\omega_{i}} / d \ln{V}$, for a large set of 
vibrational lattice frequencies, $\lbrace \omega_{i} \rbrace$ (the same than employed for the calculation of accurate 
QH free energies), for the ${\cal R}$ and ${\cal T}$ phases. The PDOS of BFCO$_{0.5}$ generally is characterized 
by a low-$\omega$ non-polar phonon region governed by Bi displacements ($0 < \omega \lesssim 4$~THz), followed by a 
medium-$\omega$ polar phonon interval dominated by transition-metal and oxygen ions ($4 \lesssim \omega \lesssim 
10$~THz), and a high-$\omega$ non-polar phonon region governed almost exclusively by oxygen atoms ($\omega \gtrsim 
10$~THz) (Fig.\ref{fig4} and Supplementary Fig.5). Since here we are interested in phase transitions occurring near 
room temperature, only those phonon excitations in the frequency interval $0 < \omega \lesssim 6$~THz are relevant 
(i.e., $k_{B}T_{\rm room} = 6.25$~THz and $\hbar \omega_{i} < k_{B}T_{\rm room}$ contribute the most to $F_{\rm vib}$ 
\cite{gould19}). The number of Bi-dominated low-energy phonon modes is higher in the ${\cal R}$ phase than in the 
${\cal T}$ phase (see PDOS peaks appearing at $\omega \approx 2$~THz in Fig.\ref{fig4}), hence the positive sign of the 
$\Delta F$ energy difference (Fig.\ref{fig3}d). However, the number of vibrational states with frequencies $2 \lesssim 
\omega \lesssim 6$~THz is larger in the ${\cal T}$ phase than in the ${\cal R}$ phase (e.g., ``bell''-like lattice modes involving concerted transition metal and oxygen displacements are missing in the latter phase --Fig.\ref{fig4}--) and consequently as the temperature 
is increased $\Delta F$ gets reduced, leading to $\Delta S > 0$. Meanwhile, positive (negative) $\gamma$ values 
indicate vibrational phonon frequencies that become ``stiffer'' (``softer'') under pressure since the bulk modulus 
of BFCO$_{0.5}$ is positive (as it occurs normally and we have explicitly checked). Consequently, based on the insets of 
Fig.\ref{fig4}, upon compression the number of phonon frequencies in the interval $2 \lesssim \omega \lesssim 6$~THz 
is further depleted in the ${\cal R}$ phase as compared to that in the ${\cal T}$ phase (that is, $-10 \lesssim \gamma_{\cal R} 
\lesssim +50$ while $-2 \lesssim \gamma_{\cal T} \lesssim +1$). This last outcome explains the fact that the stability 
$P$ span of the ${\cal T}$ phase becomes wider as the temperature is increased, leading to the positive slope  
observed in the ${\cal T}$--${\cal R}$ phase boundary (Fig.\ref{fig3}a).

The unique entropically driven room-temperature stabilization of the ${\cal T}$ phase unravelled in this study for 
BFCO$_{0.5}$ occurs at a pressure of $\sim 1$~GPa. For practical applications, it would be desirable that such a 
transformation was available at smaller compressions. Our DFT calculations carried out for BFCO$_{0.25}$ indicate that 
the critical pressure associated with the ${\cal T} \to {\cal R}$ phase transition, $P_{c}$, can be lowered drastically 
by means of composition. In particular, we estimate that $P_{c}$ may be reduced by a staggering $70$\% by increasing the 
content of Fe in the solid solution from $50$ up to $75$\% (Supplementary Fig.6). Interestingly, by repeating the same
first-principles configurational analysis than performed for BFCO$_{0.5}$, we found that both the ${\cal R}$ and ${\cal T}$ 
phases of bulk BFCO$_{0.25}$ are also ferrimagnetic and exhibit a considerable net magnetization of $0.13$~$\mu_{B}$/f.u. 
near room temperature (Supplementary Figs.7--8). Moreover, the electric polarization of bulk BFCO$_{0.25}$ changes from 
$135$ to $50$~$\mu$C/cm$^{2}$ during the ${\cal T} \to {\cal R}$ phase transition (Supplementary Fig.8), which is very 
similar to the $\Delta p$ shift estimated for BFCO$_{0.5}$. Therefore, we may conclude that the main characteristics 
of the transition involving the ${\cal R}$ and ${\cal T}$ phases in BFCO$_{0.5}$ at moderate pressures can be preserved 
and shifted down to practically ambient conditions by adjusting the relative content of Co--Fe cations in the solid solution. 
It is worth mentioning that the experimental BFCO phase diagram obtained by Azuma \emph{et al.} as a function of temperature 
and composition appears to be consistent with our theoretical findings \cite{azuma08,hojo18}. In view of the low-cost and 
scalable chemical solution methods that are available for the synthesis of BFCO solid solutions \cite{machado19}, we expect 
that our theoretical work will stimulate new and exciting experimental research on ${\cal T}$ multiferroics. 

In conclusion, we predict that a ferrimagnetic ${\cal T}$ phase can be stabilized under increasing $T$ and $P$ in bulk BFCO 
solid solutions near room temperature. This unusual $T$-induced phase transition involves a colossal increase in the electric 
polarization of $\sim 200$\% and a volume expansion of $\sim 10$\%, which is of great potential for nanoelectronics and 
energy conversion applications. Polar phonon excitations involving mixed transition-metal and oxygen ion displacements play 
a decisive role on such an extraordinary phase transformation, which is neither observed nor predicted for any of the end 
members of the solid solution. Furthermore, the thermodynamic conditions at which the entropy-driven stabilization of the 
${\cal T}$ phase occurs can be controlled adequately by varying the relative content of Co--Fe cations in the solid solution.

\section*{ACKNOWLEDGEMENTS}
Computational resources and technical assistance were provided by the Australian Government
and the Government of Western Australia through the National Computational Infrastructure
(NCI) and Magnus under the National Computational Merit Allocation Scheme and The Pawsey
Supercomputing Centre.

\end{document}